\documentclass[journal]{IEEEtran}
\usepackage{times}
\usepackage{epsfig}
\usepackage{graphicx}
\usepackage{amsmath}
\usepackage{amssymb}
\usepackage[ruled]{algorithm2e}
\usepackage{algpseudocode}
\usepackage{caption}
\usepackage{subfigure}
\usepackage{booktabs}
\usepackage{tabularx}
\usepackage{xcolor}
\usepackage{multirow}

\newcolumntype{Y}{>{\centering\arraybackslash}X}

\begin{document}
%

\title{EfficientFi: Towards Large-Scale Lightweight WiFi Sensing via CSI Compression}


\author{Jianfei~Yang,
	Xinyan~Chen,
	Han~Zou,
	Dazhuo~Wang,
 	Qianwen~Xu,
 	and~Lihua~Xie,~\IEEEmembership{Fellow,~IEEE}
 
 \thanks{
	J. Yang, X. Chen, D. Wang and L. Xie are with the School of Electrical and Electronics Engineering, Nanyang Technological University, Singapore (e-mail: yang0478@ntu.edu.sg; elhxie@ntu.edu.sg).
 
 	H. Zou is the corresponding author, and he is with the Department of Electrical Engineering and Computer Sciences, University of California, Berkeley, USA (e-mail: enthalpyzou@gmail.com).
 
	Q. Xu is with the Department of Electric Power and Energy Systems, KTH Royal Institute of Technology, Sweden (e-mail: qianwenx@kth.se).
	
	This work is supported by NTU Presidential Postdoctoral Fellowship, ``Adaptive Multimodal Learning for Robust Sensing and Recognition in Smart Cities'' project fund, in Nanyang Technological University, Singapore.
	}
}

\markboth{}%
{Shell \MakeLowercase{\textit{et al.}}: Bare Demo of IEEEtran.cls for IEEE Journals}

\maketitle

\begin{abstract}
   WiFi technology has been applied to various places due to the increasing requirement of high-speed Internet access. Recently, besides network services, WiFi sensing is appealing in smart homes since it is device-free, cost-effective and privacy-preserving. Though numerous WiFi sensing methods have been developed, most of them only consider single smart home scenario. Without the connection of powerful cloud server and massive users, large-scale WiFi sensing is still difficult. In this paper, we firstly analyze and summarize these obstacles, and propose an efficient large-scale WiFi sensing framework, namely EfficientFi. The EfficientFi works with edge computing at WiFi APs and cloud computing at center servers. It consists of a novel deep neural network that can compress fine-grained WiFi Channel State Information (CSI) at edge, restore CSI at cloud, and perform sensing tasks simultaneously. A quantized auto-encoder and a joint classifier are designed to achieve these goals in an end-to-end fashion. To the best of our knowledge, the EfficientFi is the first IoT-cloud-enabled WiFi sensing framework that significantly reduces communication overhead while realizing sensing tasks accurately. We utilized human activity recognition and identification via WiFi sensing as two case studies, and conduct extensive experiments to evaluate the EfficientFi. The results show that it compresses CSI data from 1.368Mb/s to 0.768Kb/s with extremely low error of data reconstruction and achieves over 98\% accuracy for human activity recognition. 
\end{abstract}

\begin{IEEEkeywords}
Channel state information; WiFi-based sensing; multi-task learning; variational auto-encoder; discrete representation learning; deep neural network.
\end{IEEEkeywords}


\section{Introduction}
\IEEEPARstart{W}{iFi} technology has been ubiquitous with the rapid growth of wireless IoT devices and the increasing demands of internet access. To improve throughput for higher demands of wireless data traffic, Multiple-Input Multiple-Output (MIMO) technology is proposed along with Orthogonal
Frequency-Division Multiplexing (OFDM) \cite{love2003grassmannian}. For each antenna pair between receiver and transmitter devices, MIMO provides Channel State Information (CSI) at each subcarrier frequency to describe the propagation situations of wireless signals. Recently, apart from the communication purpose, CSI measurements are leveraged for sensing purposes \cite{yang2018device}. For example, by signal modeling or deep neural networks, CSI patterns for human activities can be extracted, enabling accurate human activity recognition at smart homes or buildings \cite{yang2019learning,zou2017freedetector,zou2018deepsense,zou2018identification}. 

\begin{figure}[t]
	\centering
	\includegraphics[width=1.0\linewidth]{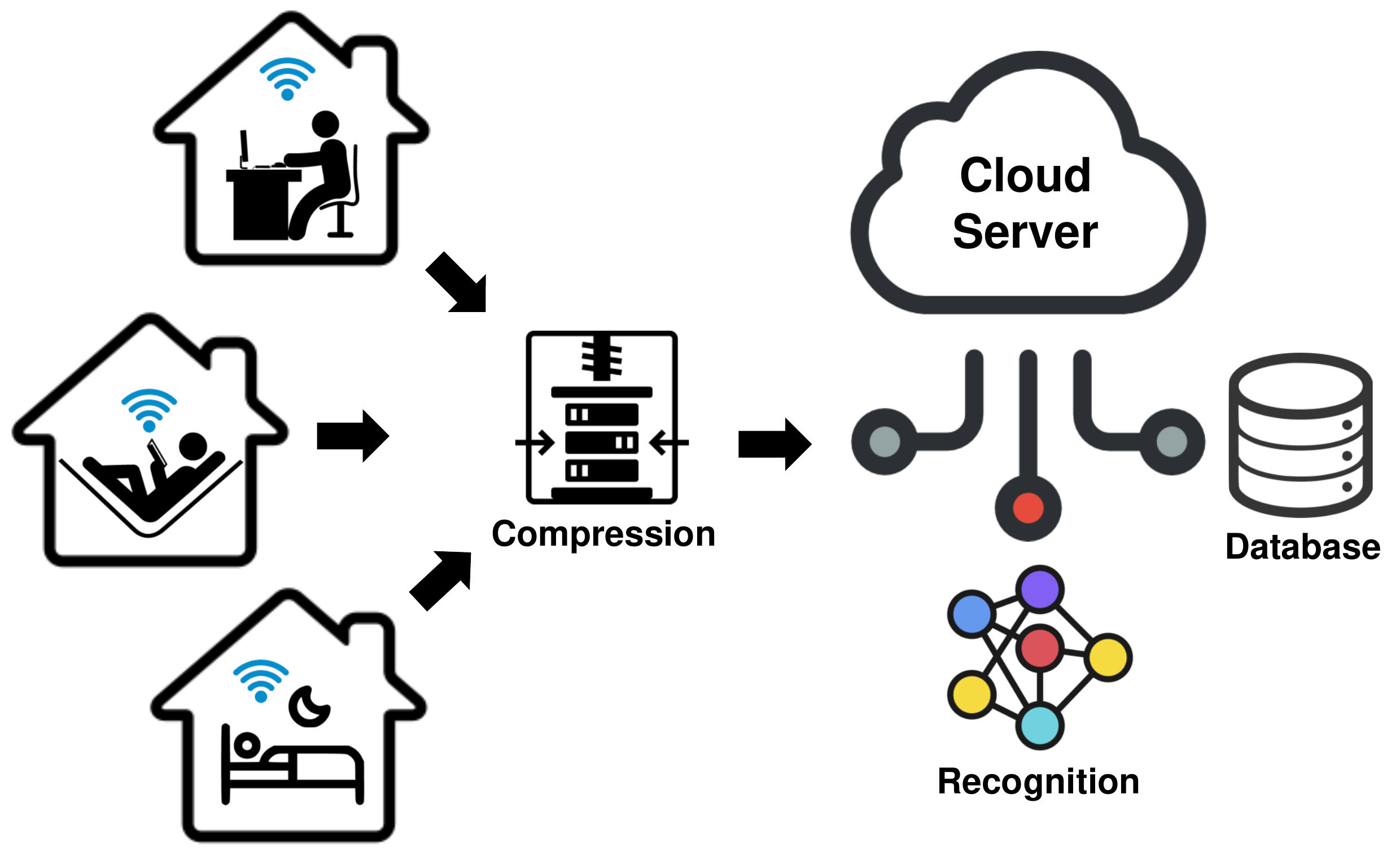}
	\caption{The illustration of the proposed EfficientFi framework. The EfficientFi is designed for three functions: data compression for efficient transmission, data restoration for database logging, and recognition task performing. It addresses the limitations of large-scale WiFi sensing compared to existing single-user systems.}
	\label{fig:intuition}
\end{figure}

Compared to other smart sensing solutions, WiFi sensing has various merits. Firstly, WiFi sensing is quite cost-effective due to the reuse of existing WiFi infrastructures for wireless communications. Secondly, unlike the wearable sensor based solutions \cite{chen2017robust}, WiFi sensing is non-intrusive and thus convenient to use. Thirdly, though camera-based solutions provide high sensing granularity, they suffer from occlusion or poor illumination at night \cite{xu2021arid}, and often arouse privacy issue, which can be overcome by WiFi sensing. Having these advantages, WiFi sensing has gained much attention in ubiquitous computing research. Based on the strong recognition capacity of deep neural networks, many WiFi sensing applications have been developed, such as activity recognition, gesture recognition and person identification.

Although there have been masses of works on robust WiFi sensing systems with various recognition methods (e.g. model-based and learning-based algorithms), current solutions still encounter serious problems when they are deployed in the real-world scenarios. The main challenge lies in the limited computation resources on the edge side \cite{shi2016edge}, considering the requirement of low cost and low power for users. The solution is to transmit the CSI data to a cloud server that conducts centralized computing for these data and returns the results to users. However, as CSI measurement has high dimension and sampling rate for sensing, real-time CSI stream can lead to severe communication burden that may directly hinder the basic functionality of WiFi, i.e. Internet access. Therefore, for transmission of large-scale WiFi sensing data from user to cloud center, a compression algorithm is highly demanded.

Supposing that the compressed CSI data has been transmitted to cloud servers for recognition, we still encounter new problems at the cloud server side. Not only an accurate sensing model is required at cloud servers, but also the transmitted CSI data should be reconstructed for data logging requirement of Internet of Things (IoT) systems and the incremental learning of the recognition model \cite{ray2016survey}, which have been proved to be demanded in healthcare \cite{hassanalieragh2015health} and other IoT research fields \cite{singh2020iot}. To this end, two significant objectives, an accurate CSI-based recognition model and the reconstruction of CSI data, should be achieved simultaneously. In this manner, WiFi sensing can be accurate, communication-efficient, and data-traceable, which forms an IoT-enabled cost-effective sensing system for large-scale use cases, as shown in Figure \ref{fig:intuition}.

To deal with the aforementioned challenges, we propose EfficientFi that has three functions: CSI compression, CSI restoration and CSI-based recognition, which bridges the gap between existing single-user sensing system and large-scale WiFi sensing. It is composed of a novel quantized feature learning algorithm and a consensus learning framework for training accurate recognition model simultaneously. Specifically, the feature extraction and classification functions are assigned to the edge and cloud side, respectively. In the WiFi router, robust CSI feature is extracted and also compressed for transmission. In the cloud server, the feature is fed into a classifier for recognition purpose and a decoder for CSI data reconstruction. Though the inference model is easy to design, it is a nontrivial task on how to quantize the deep representation, reconstruct the CSI data and train the compression model with the recognition jointly. To this end, we propose a multi-task learning framework that enables the main functionalities of EfficientFi. The EfficientFi model is an end-to-end model that can be optimized offline in the server side. Once it has been trained, it has the capacity of quantizing any CSI feature by a codebook that is employed in both edge and cloud side. Then we deploy the feature extractor in the edge side, and preserve the classifier and the decoder on the cloud side. In this manner, EfficientFi is able to transmit compressed CSI feature to the cloud server for recognition efficiently. Simultaneously, the CSI can be restored for further usage. To the best of our knowledge, EfficientFi is the first work that considers the challenges of the large-scale WiFi sensing scenario and deals with these problems.

We summarize the main contributions as follows:
\begin{itemize}
	\item We firstly analyze and identify the communication limitation of current large-scale WiFi sensing solutions based on the edge-cloud computation architecture.
	\item To address the challenge, we propose EfficientFi, a novel framework that compresses CSI into a quantized low-dimensional space at the edge side and decodes compressed CSI at cloud server for low-cost communication.
	\item We further develop a joint learning strategy to enforce the compressed space to be discriminative for CSI-based recognition tasks. The EfficientFi is the first system that is able to compress and further identify the CSI profiles for large-scale wireless human sensing applications.
	\item To validate the EfficientFi, we conduct real-world experiments including WiFi human activity recognition and WiFi human identification. The results show that the EfficientFi reduces the communication burdens by over 1,700 times without degrading recognition performance.
\end{itemize}


\section{Related Work}\label{sec:related-work}
In this section, we discuss the relevant work in large-scale WiFi sensing systems: existing WiFi sensing applications, deep learning models for WiFi sensing and CSI compression in communication field.

\subsection{WiFi Sensing Systems and Applications}
WiFi-based sensing leverages Channel State Information (CSI) between a transmitter and a receiver for human motion detection and recognition \cite{zhang2020device}. Compared to Inertial Measurement Unit (IMU) and camera-based sensing, WiFi sensing is device-free, cost-effective and privacy-preserving, which is an appealing sensing solution for smart home and buildings. The first tool that enables WiFi sensing is the Intel 5300 Network Card that collects 30 subcarriers CSI data \cite{halperin2011tool}. Then the Atheros tool \cite{xie2015precise} is proposed to extract 56 subcarriers CSI data with 20MHz bandwidth. A new platform is constructed based on the Atheros tool, which enables CSI extraction of 114 subcarriers with 40MHz directly on IoT devices \cite{yang2018device}. This new tool improves the CSI resolution and enables CSI extraction from Atheros chip at WiFi Access Point (AP) instead of mini-PC for Intel 5300. Many WiFi sensing systems have been proposed for various applications including occupancy detection \cite{zou2017freedetector}, crowd counting \cite{zou2018device}, human activity recognition \cite{zou2018deepsense,zou2017multiple,zou2019wifi}, respiration monitoring \cite{liu2021wiphone}, person identification \cite{zou2018identification} and gesture recognition \cite{zou2018robust,yang2019learning}. Recently, CSI can be extracted between a smartphone and a WiFi AP by Shadow Wi-Fi \cite{schulz2018shadow}, which further increases the application range of WiFi sensing. Since every smartphone is equipped with a WiFi module, more CSI data will be extracted if this tool is widely applied. WiFi-based applications will be promoted to large-scale user scenario. In summary, existing WiFi sensing systems only consider the single or multiple user scenario,  but they cannot empower large-scale WiFi sensing due to the communication cost, which can be overcome by the proposed EfficientFi.

\subsection{Deep Learning Models for WiFi Sensing}
When it comes to the recognition model of WiFi CSI data, model-based methods and learning-based methods are two main streams. Model-based methods leverage radar theories and signal processing techniques to model the scenario when persons affect WiFi propagation, such as Fresnel Zone model \cite{zhang2021fresnel}. Learning-based models are data-driven, which leverages statistical features or deep neural networks by pattern learning and optimization, which is more relevant to our paper. Compared to model-based methods, deep learning models can formulate and learn more complicated patterns, but require massive data collection \cite{ma2019wifi}. Chen et al. propose to extract temporal CSI patterns via LSTM for human activity recognition \cite{chen2018wifi}. Zou et al. propose DeepSense that integrates temporal and spatial features for CSI-based activity recognition. Yang et al. propose to recognize gestures from CSI by a Siamese recurrent convolutional network \cite{yang2019learning}. Recurrent neural networks are also leveraged for human identification \cite{choi2017deep}. Also, deep learning approach shows better performance than traditional machine learning methods for CSI-based indoor localization \cite{wang2016csi}. Deep learning models significantly improve the recognition granularity of WiFi sensing, but require more computation overhead and thus cloud server is normally demanded. The proposed EfficientFi is a novel framework with deep CSI compression, quantization and recognition.

\subsection{CSI Compression in Communication}
In 5G communication, the massive multiple-input multiple-output (MIMO) system has been widely adopted as a premium technology. By exploiting CSI at base stations, the MIMO system can significantly reduce multi-user interference. To this end, the CSI is acquired at the user equipment and then transmitted back to the base station via feedback links \cite{larsson2014massive}. The task of CSI feedback in MIMO system is quite similar to our task for the CSI compression part, though our objectives and scenarios are totally different. To encode and decode CSI in a MIMO system, numerous studies have been proposed to reduce feedback overhead. Classic methods leverage compressive sensing \cite{donoho2009message} or LASSO $l_1$-solver \cite{daubechies2004iterative}. However, the simple prior cannot perfectly recover compressive CSI since the channel matrix is approximately sparse. Though advanced algorithms impose elaborate priors on reconstruction \cite{metzler2016denoising}, the hand-crafted priors still hinder the recovery quality. Then CSINet \cite{wen2018deep} comes into existence, which learns an auto-encoder from training samples, substantially boosting CSI recovery quality. Then a revamped method achieves better trade-off between CSI quality and feedback overhead \cite{mashhadi2020distributed}. Different from CSI feedback research in communication field, CSI-based sensing technology requires both a compressed (i.e. codewords) and discriminative feature space for sensing and recognition applications. Therefore, the existing methods mentioned above are not applicable since their CSI codewords are only prepared for compression and communication. To the best of our knowledge, EfficientFi is the first work that learns a compressed discriminative feature space for WiFi sensing, which aims to empower large-scale wireless sensing techniques for massive users.

\begin{figure*}[h]
	\centering
	\includegraphics[width=0.9\textwidth]{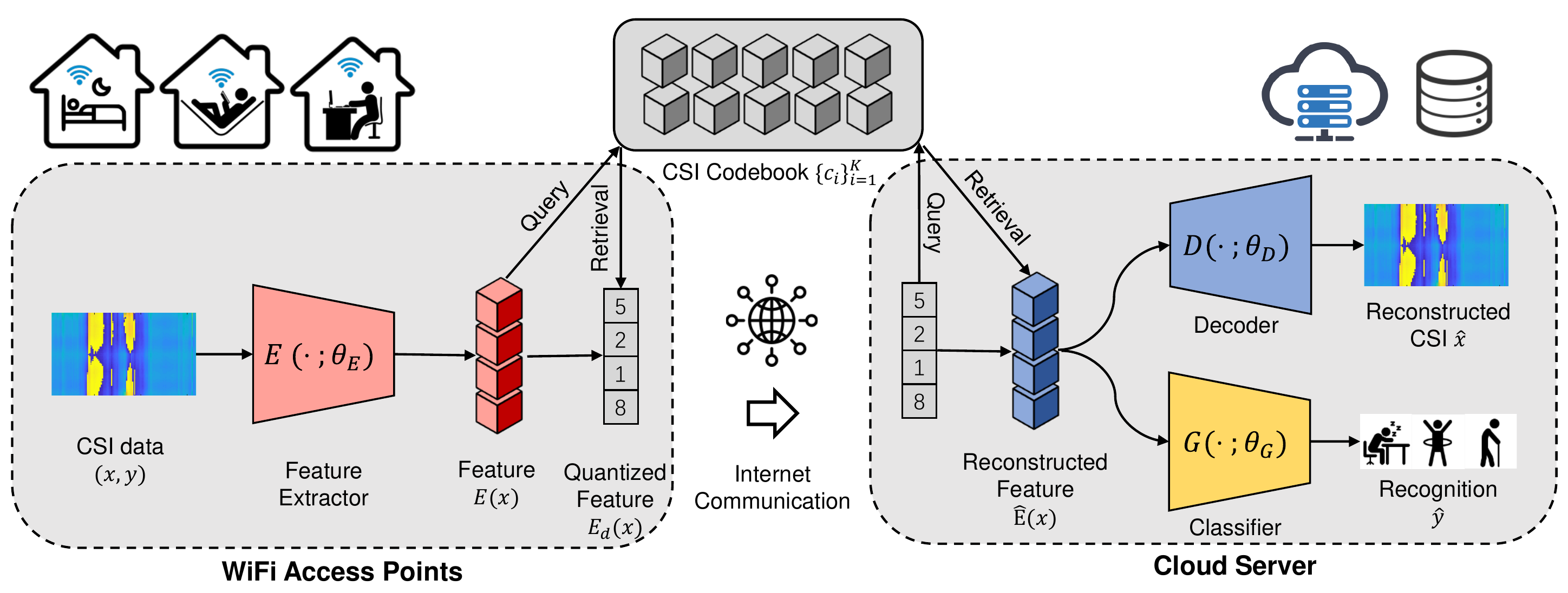}
	\caption{The design of the proposed EfficientFi framework.}
	\label{fig:framework}
\end{figure*}

\section{Preliminaries}\label{sec:preliminaries}
\subsection{Channel State Information}
In wireless communication, channel state information describes the channel properties of a communication link, and thus reflects how wireless signals propagate in a physical environment where reflections, diffractions and scattering may happen. Modern WiFi devices leverage Orthogonal Frequency Division Multiplexing (OFDM) at the physical layer and follow IEEE 802.11 standard. Multiple antennas are available for transmitter and receiver devices. For each pair of two antennas, CSI records wireless signal characteristics including time delay, amplitude attenuation, and phase shift of multi-paths on each communication subcarrier. Hence, CSI has higher granularity when it is compared to received signal strength. The WiFi signals can be modeled as Channel Impulse Response (CIR) $h(\tau)$ in frequency domain:
\begin{equation}
h(\tau)=\sum_{l=1}^{L}\alpha_l e^{j\phi_l} \delta(\tau-\tau_l),
\end{equation}
where $\alpha_l$ and $\phi_l$ represent the amplitude and phase of the $l$-th multi-path component respectively, $\tau_l$ is the time delay, $L$ indicates the total number of multi-path and $\delta(\tau)$ denotes the Dirac delta function. In the real-world situations, based on specific tool \cite{xie2015precise}, the OFDM receiver is able to sample the signal spectrum at subcarrier level, which comprises amplitude attenuation and phase shift via complex number. These estimation can be represented by:
\begin{equation}
H_i=||H_i||e^{j \angle H_i}
\end{equation}
where $||H_i||$ and $\angle H_i$ are the amplitude and phase of $i$-th subcarrier, respectively. The normal CSI compression for communication purpose aims to estimate this matrix for each packet. In WiFi sensing, it is not necessary to estimate the whole matrix but its part. For example, human activity recognition, the amplitude $||H_i||$ should be compressed. The compression process encodes the data to a low-dimensional space and then the decoder restores the data to its original space with small reconstruction errors. In our scenario, we further expect such compressed low-dimensional space to be discrete, so that it can be transmitted with lower communication resources.

\subsection{WiFi Sensing and Limitations}
Previous works have shown that CSI patterns are quite different when persons perform different activities. The reason is that human activities can affect the propagation of wireless signals and this phenomenon can be captured by CSI data \cite{ma2019wifi}. Collecting CSI at the receiver side, machine learning or model-based analytic methods can extract these CSI patterns and thus enable many applications. Furthermore, one transmitter and multiple receivers can form more links, which provides abundant CSI data for multi-user applications \cite{tan2019multitrack}. Transfer learning methods further enable the learning model to have strong generalization ability across sites and time \cite{zou2018robust, zhang2018crosssense}. These methods only regard WiFi sensing as a local sensor network with local computation resources. Nevertheless, in real-world large-scale WiFi sensing, thousands of CSI data will be transmitted to a cloud server for computation and data record, which is similar to the current visual recognition services on cloud \cite{agrawal2015cloudcv}. This will arouse new challenges to WiFi sensing: how to reduce communication cost and carry out model inference simultaneously in cloud servers. For example, if a pair of WiFi sensing devices operates on 40MHz (i.e. 114 subcarriers) with three pairs of antennas and the sampling rate is 500Hz for CSI collection, then the communication cost will be $3\times 114\times 500 \times 2 \times 4$ bytes/s (i.e. 1.368Mb/s). In this paper, the EfficientFi learns a discriminative compressed feature space for CSI, bridging the gap between edge and cloud computing for large-scale WiFi sensing.

\section{EfficientFi Design}\label{sec:design}
\subsection{System Overview}
As shown in Figure \ref{fig:framework}, the proposed EfficientFi works in a common scenario when massive CSI data is transmitted to a cloud server from many WiFi Access Points (AP) of users. The objectives of our method include efficient communication of CSI data, reconstruction of CSI data at server database and accurate recognition tasks based on CSI. At WiFi AP side, our method firstly extracts features by a lightweight Convolutional Neural Network (CNN), and then compresses these features to a quantized vector by looking up their nearest vectors in a CSI codebook. Then at cloud server side, the quantized vector can be restored and reconstructed to the original CSI data for database record, and the reconstructed feature is fed into a classifier for specific WiFi sensing tasks, such as human activity recognition. The compression design is inspired by Vector Quantization Variational Auto-Encoder (VQ-VAE) \cite{van2017neural}. In comparison, our EfficientFi not only requires high compression rate, but also demands the capacity of recognition tasks, which brings many challenges.

For the aforementioned process, we formulate each component of our method. Given the labeled CSI data $\{x, y\}$, the novel EfficientFi consists of a feature extractor $E(\cdot;\theta_E)$, a decoder network $D(\cdot;\theta_D)$ and a task classifier $G(\cdot;\theta_G)$, parameterized by $\theta_E$, $\theta_D$ and $\theta_G$, respectively. CSI codebook $c \in \mathbb{R}^{K\times D}$ consists of $K$ $D$-dim vectors $c_i$ for quantization. Denote normal continuous feature and quantized feature as $E_c(x)$ and $E_d(x)$, respectively. Denote $z_j$ as a random variable that represents the $j$-th position of the discrete latent variables $E_d(x)$. The objective of our method is to reconstruct the original sample $x$ by $c$, $E_c(x)$ and $D$. Our algorithm firstly learns $\theta_E$ and $\theta_D$ by minimizing the global reconstruction error, and then learns the codebook. Finally, the parameters of the classifier $\theta_G$ is jointly optimized with the feature extractor $\theta_E$ so that $E$ can capture the discriminative parts for recognition.

\subsection{Discrete Representation Learning for Efficient Communication}
At WiFi AP side, when persons perform various activities, such activities can be captured by CSI data. In EfficientFi, CSI data is firstly fed into a CNN feature extractor $E(\cdot;\theta_E)$ and a discriminative feature $E_c(x)$ is obtained. The feature $E_c(x)$ is a sequence of vectors. Assume that similar vectors of these features in $E_c(x)$ can be found in our CSI codebook $c$. For each entry of $E_c(x)$, we can search its nearest neighbor in the CSI codebook and generate a new quantized feature $E_d(x)$ by the indices of these similar vectors. The posterior categorical distribution $q(z_j|x)$ probabilities for each position (entry) $j$ is given by a one-hot form as follows:
\begin{equation}\label{eq:codebook}
q(z_j|x)=\left\{
	\begin{array}{rcl}
	1 & & {\text{for }k=\arg\min_i \|E_c(x)-c_i \|_2}\\
	0 & & {\text{otherwise}}\\
	\end{array}. \right.
\end{equation}
In this manner, all the entries of such representation are integers, formed by one-hot vectors. Similarly, the reconstructed feature $\hat{E_d}(x)$ can be obtained by an opposite way at cloud server. Through the decoder $D(\cdot;\theta_D)$, we can reconstruct the CSI data for database record at cloud.

\subsection{Joint Learning of Quantized Representation and Recognition Model}
Apart from compression and restoration, we still expect the EfficientFi model to have recognition capability, which puts forward new requirements for our quantized feature: to be discriminative for specific tasks. We design a classifier $G(\cdot;\theta_G)$ made up of fully connected layers. For a specific task, such classifier aims to make predictions according to our reconstructed CSI feature $\hat{E_c}(x)$. For example, it can predict categories of human activity for WiFi-based activity recognition task. The EfficientFi can also support other applications using the same architecture but different annotations, such as gesture recognition and human identification. 

\subsection{Learning and Optimization}
Though the forward process of our proposed EfficientFi seems reasonable, two difficulties are encountered in practice: how to obtain a codebook for encoding and decoding, and how to train the whole model in an end-to-end manner. On the one hand, without a comprehensive codebook, the CSI data cannot be perfectly reconstructed because CSI can be distinct in its numeric range. On the other hand, since the generation of quantized feature $E_d(x)$ in Eq.(\ref{eq:codebook}) is non-derivative, the model cannot be directly trained by classic back-propagation algorithm \cite{lecun1988theoretical}.

To deal with these problems, VQ-VAE offers a straight forward way. To obtain a meaningful codebook, we make $c$ to be learnable parameters that are updated by back-propagation. Its learning objective is to minimize the difference between $E_c(x)$ and $E_d(x)$. For end-to-end learning, we can enforce the gradient of the decoder $\triangledown L_D$ to be identical to that of the encoder without any modification. Such straight-through gradient estimation preserves useful information to guide the encoder to obtain decreasing CSI reconstruction loss by changing its output.

Having synthesized these analyses, we design the three learning objectives of the EfficientFi which can be optimized by back-propagation in an end-to-end fashion:
\begin{equation}\label{eq:main-ed}
	\min_{\theta_E,\theta_D} \mathcal{L}_r=\| x-D(E_c(x)+\text{sg}[E_d(x)-E_c(x)]) \|^2_2,
\end{equation}
where $\text{sg}[\cdot]$ represents the stopgradient operator that outputs identity results at forward computation time and has zero partial derivatives. This loss updates both the encoder and the decoder by calculating the mean squared loss between the original sample and its estimation $D(E_x(x))$. Since the gradient cannot be directly passed between $E_c(x)$ and $E_d(x)$, we apply the straight-through estimator \cite{bengio2013estimating} between $E_c(x)$ and $E_d(x)$. Consequently, the forward pass of this loss is $\| x-D(E_c(x)+E_d(x)-E_c(x)) \|^2_2=\| x-D(E_d(x)) \|^2_2$ while the backward gradient is only given by $\| x-D(E_c(x)) \|^2_2$ without considering the content inside stopgradient operator.

The second objective is to learn the codebook as follows:
\begin{equation}\label{eq:main-c}
	\min_c \mathcal{L}_c=\|\text{sg}[E_c(x)]-E_d(x)\|^2_2.
\end{equation}
This loss aims to minimize the distance between $E_c(x)$ and $E_d(x)$ by identifying better codebook vectors $c_i$. Fixing $\theta_E$ and $\theta_D$, the codebook can be updated by the dictionary learning algorithm that learns to move the codebook vectors to the encoder output $E_c(x)$ by the $l_2$ error, which is namely the Vector Quantization (VQ) algorithm \cite{gray1984vector}.

The third objective simultaneously updates the feature extractor and the classifier. It aims to learn a compressed feature space that can also serve for WiFi sensing tasks, which can be formulated as
\begin{equation}\label{eq:main-eg}
	\min_{\theta_E, \theta_G} \mathcal{L}_e=\lambda \| E_c(x)-\text{sg}[E_d(x)] \|^2_2+\mathcal{L}_y(x,y),
\end{equation}
where $\lambda$ is a hyper-parameter that achieves the trade-off between the training of auto-encoder and the classifier, and $\mathcal{L}_y$ is the cross-entropy loss for a specific $T$-way classification task for WiFi sensing applications:
\begin{equation}
\mathcal{L}_{y}(x,y)=-\mathbb{E}_{(x,y)}\sum_t \big[ \mathbb{I}[y=t] \log \big(\sigma(G(\hat{E_c}(x))) \big) \big],
\end{equation}
where $\sigma$ is the softmax function, and $\mathbb{I}[y=t]$ means a 0-1 function. For the correct category $t$, such term is equal to one. Here we fix the decoder and train the encoder to mimic the discrete feature $E_d(x)$. This term also ensures that the encoder keeps reasonable updating speed with the codebook learning. In addition, to make the compressed feature space discriminative, we integrate this commitment loss with the classification loss, and formulate a multi-task learning objective.

\subsection{Algorithm Summary}
The overall learning objective of the EfficientFi is written as
\begin{equation}\label{eq:total}
	\min_{\theta_E,\theta_D, \theta_G,c}\mathcal{L}_{EfficientFi}=\mathcal{L}_r+\mathcal{L}_c+\mathcal{L}_e.
\end{equation}
Note that the overall loss only shows all of the loss functions and parameters to be optimized in EfficientFi for better understanding. Each optimization process is conducted separately. To be specific, we illustrate the training and deployment of the EfficientFi in Algorithm \ref{algorithm}. We firstly train the EfficientFi model offline using existing CSI data. Then we deploy the feature extractor $E$ at WiFi AP and equip cloud server with the decoder $D$ and the classifier $G$. The well-trained CSI codebook is stored in both WiFi APs and cloud server for compression and restoration.

The EfficientFi firstly deals with the large-scale WiFi sensing and can achieve better performance than existing CSI compression methods due to two reasons. The first reason is how the codebook is learned and then serves for the encoder and the decoder. The learnable parameters in the codebook can better capture the CSI patterns and conduct the reconstruction, when compared to CSINet that separately proceed the quantization. This is also demonstrated in the computer vision field~\cite{van2017neural}. The second merit of the EfficientFi comes from the multi-task learning scheme, where the feature learning integrates the recognition and reconstruction objectives and thus captures more semantic features.

\begin{algorithm}
	\caption{Training and Deployment of EfficientFi}\label{algorithm}
	\KwIn{labelled CSI data $(x,y)$, weight parameters $\theta_E,\theta_G,\theta_D$, the CSI codebook $c$, the total number of iterations $N_{itr}$}
	\BlankLine
	{Initialize the feature extractor $E$, the decoder $D$ and the classifier $G$ by $\theta_E,\theta_D,\theta_G$, respectively}\;
	
	\While{$num\_iteration \leqslant N_{itr}$ }
	{1. Input the labelled data $(x,y)$\;
	2. Extract the feature $E_c(x)$ and $E_d(x)$ by query $c$\;
	3. Restore the reconstructed feature $\hat{E_c}(x)$ \;
	4. Reconstruct CSI data $\hat{x}$ by $D(\cdot)$\;
	5. Predict the sensing results by $G(\cdot)$\;
	6. Update $\theta_E,\theta_D$ by optimizing Eq.(\ref{eq:main-ed})\;
	7. Update the codebook by minimizing Eq.(\ref{eq:main-c})\;
	8. Update $\theta_E,\theta_G$ by minimizing Eq.(\ref{eq:main-eg})\;
	}
	
	{Deploy $E$ at edge side, $D$ and $G$ at cloud server};\
	
	{Store the codebook $c$ at both sides};\
	
\end{algorithm}

\begin{table*}[htp]
	\centering
	\begin{tabular}{c|c|c|c}
		\toprule \midrule
		Layer Index & Feature Extractor $E$            & Decoder Network $D$         & Label Predictor $G$   \\ \midrule
		input           & \multicolumn{3}{c}{CSI data: 3 $\times$ 114 $\times$ 500 (antenna pairs $\times$ subcarrier $\times$ sampling rate)}                         \\ \midrule
		1           & Conv 32$\times$(15,23), stride 9, ReLU      & Max-unpool (1,2), stride (1,2)         & 128 dense \\ \midrule
		2           & Conv 32$\times$(3,7), stride 1, ReLU &   ConvTrans 96$\times$(3,7), stride 1, ReLU    &   6 dense, softmax                \\ \midrule
		3           & Max-pool (1,2), stride (1,2)  & ConvTrans 64$\times$(3,7), stride 1, ReLU &                   \\ \midrule
		4           & Conv 64$\times$(3,7), stride 1, ReLU & Max-unpool (1,2), stride (1,2)   &                   \\ \midrule
		5           & Conv 96$\times$(3,7), stride 1, ReLU & ConvTrans 32$\times$(3,7), stride 1, ReLU &        \\ \midrule
		6           & Max-pool (1,2), stride (1,2)  & ConvTrans 32$\times$(15,23), stride 9, ReLU & \\ \midrule \bottomrule
	\end{tabular}
	\caption{The network architecture used in the EfficientFi experiments. For Conv A$\times$(H,W), A denotes the channel number, and (H,W) represents the height and width of the operation kernel. This applies to Convolution (Conv), ConvTranspose (ConvTrans), Max-pooling (Max-pool) and Max-unpooling (Max-unpool) layers.}\label{table:network}
\end{table*}

\section{Experiment}\label{sec:experiment}
\subsection{Experimental Setup}
To demonstrate the effectiveness of the EfficientFi, we choose CSI-based Human Activity Recognition (HAR) and Human Identification (Human-ID) as the case studies, and implement the EfficientFi between WiFi AP and a server. Then we conduct extensive experiments to evaluate the performance of EfficientFi with respect to compression rate, data restoration quality, and the recognition accuracy of tasks.

\textbf{System Design.} To fully evaluate the compression functionality, we aim to use as many subcarriers as possible, increasing the compression difficulty. To this end, we implement our system on two TP-Link N750 APs that serve as a transmitter and a receiver. They have three antenna pairs, operating with 40Mhz bandwidth under 5GHz, which allows us to extract complete 114 subcarriers of CSI data at each timestamp. The cloud server is simulated by a local powerful server equipped with one NVIDIA RTX 2080Ti GPU. The system is similar to \cite{yang2018device}. Compared to classic Intel 5300 NIC \cite{halperin2011tool} that only extracts 30 subcarriers of CSI for each pair of antennas, more subcarriers render our scenario to be more challenging since our EfficientFi is required to reconstruct more CSI dimensions.

\begin{figure}[t]
	\centering
	\subfigure[Layout]{\includegraphics[width=0.7\linewidth, angle=0]{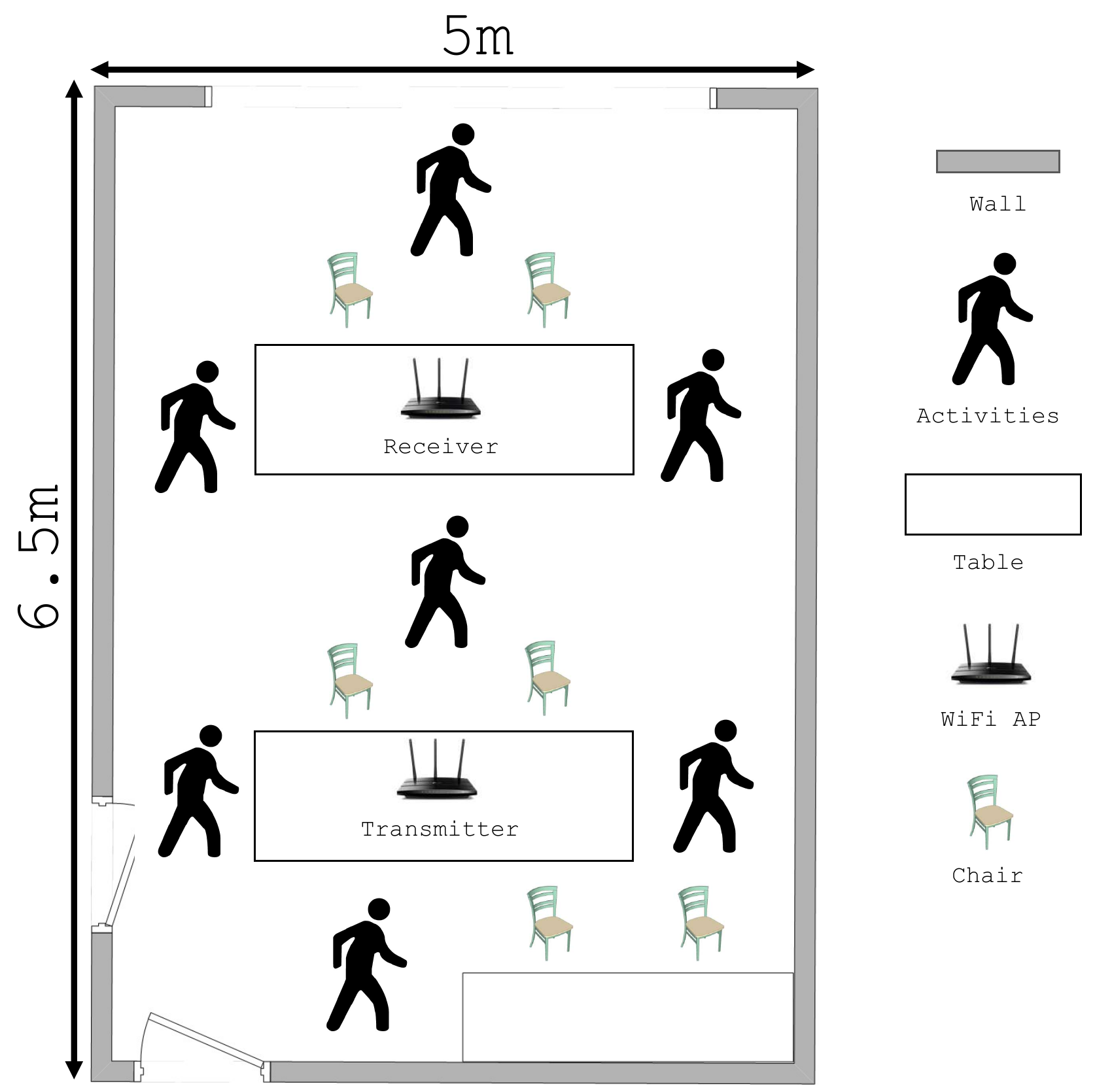}}
	\subfigure[Real scene]{\includegraphics[width=0.7\linewidth, angle=0]{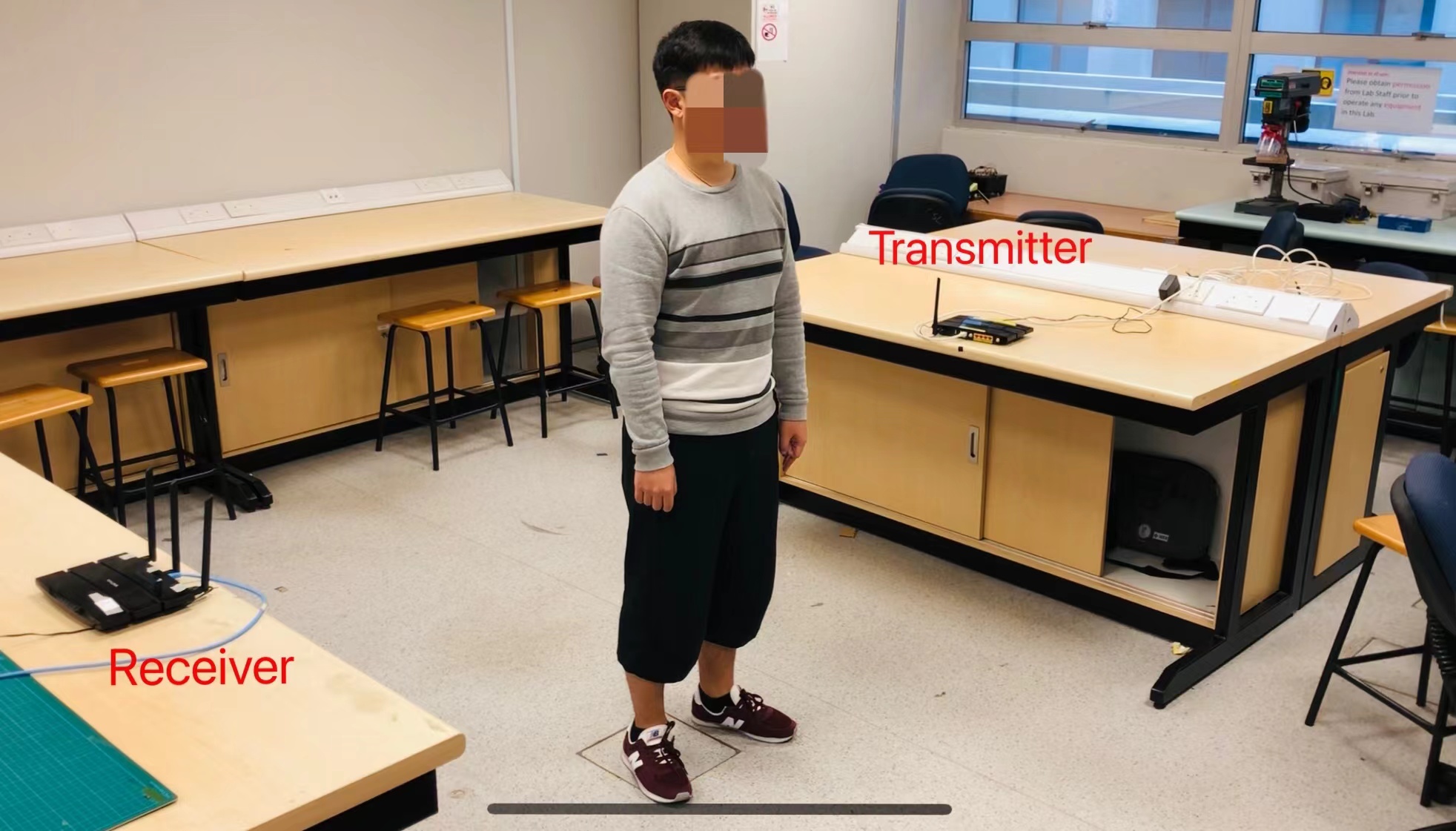}}
	\caption{The experiment environment in the lab.}
	\label{fig:layout}
\end{figure}

\textbf{Data Collection.} We collect the data at a sampling rate of 500Hz, and thus the number of data samples per second is $3\times 114\times 500$. For human activity recognition, we collect six human activities including running, walking, falling down, boxing, circling arms, and cleaning floor for 20 subjects (13 males and 7 females). Each subject performs 20 times for each activity, and thus every category has 400 samples. The environment for data collection is shown in Figure \ref{fig:layout}. For person identification, the subjects are demanded to walk through the Line-of-Sight (LOS) of the pair of WiFi APs from different directions. The CSI profiles of 15 subjects are recorded in our dataset, the number of samples for each subject is 60. All categories of human activities and subjects are utilized for evaluation in the following experiments. As our mission is to test the compression error and the possible accuracy drop after compression, the testing scenario does not need to be very complicated. We split the dataset to training and testing set by 8:2. Only the amplitude is employed for compression and recognition in our experiment, and the CSI data is not pre-processed by any denoising scheme.

\textbf{Network Implementation.} The network design of $E,D,G$ has been shown in Table \ref{table:network}. Note that deeper networks could be utilized for more difficult tasks. Briefly speaking, $E$ and $D$ consist of four convolution layers or convolution transpose layers that are computation-efficient. The classifier consists of two fully connected layers with a softmax function. The embedding dimension $D$ of the codebook $c$ is 256, and the code number $K$ varies among $[64, 128, 256, 512, 1024]$. The weight $\lambda$ is set to 0.5. These hyper-parameters are analyzed by empirical studies in Section \ref{sec:hyper}, where we choose the best hyper-parameters. We train the model for 100 epochs with a batch size of 128. The two classifiers for HAR and Human-ID are trained separately. The EfficientFi is optimized by a mini-batch SGD with an initial learning rate of 0.01 and a momentum of 0.9. The learning rate is decayed by 0.1 at 40th and 80th epochs. The whole training process only takes 10 mins in our server.

\textbf{Criterion.} For evaluation, we mainly compare the compression rate, CSI restoration quality and classifier performance by the communication cost per second, reconstruction error and the accuracy. The original communication cost is 4 bytes float for $3\times 114\times 500$ amplitude data, leading to 1.368Mb/s. The compression rate $\gamma$ in our paper is the ratio of 1.368Mb/s and the test communication cost. For restoration quality, we adopt Normalized Mean Squared Error (NMSE) that is quantified by decibel (dB):
\begin{equation}
	\text{NMSE}=\mathbb{E}\bigg[\frac{\|x-\hat{x}\|^2_2}{\|x\|^2_2} \bigg].
\end{equation}
The recognition accuracy is denoted as the ratio of true predicted samples and all testing samples.

\textbf{Baselines.} For CSI compression, we compare our method with three state-of-the-art methods. The vanilla LASSO $l_1$-solver \cite{daubechies2004iterative} is the basic optimization algorithm that considers the simplest sparsity prior. Then BM3D-AMP \cite{mashhadi2020distributed} shows the best compressive recovery performance in image reconstruction. CSINet \cite{wen2018deep} leverages asymmetric deep auto-encoder that achieves remarkable performance. We also compare EfficientFi with the vanilla auto-encoder network though it cannot be utilized due to the unavailability of quantization. For human activity recognition, E-eyes \cite{e-eyes} and CARM \cite{carm} are two baselines that leverage statistical features of CSI patterns. For gait recognition, we choose WiWho \cite{wiwho} and AutoID \cite{zou2018identification} as two baseline methods.

\subsection{Evaluation on HAR and Human-ID}
We firstly compare the EfficientFi with existing state-of-the-art methods for the human activity recognition task and the person gait identification task in Table \ref{tb:overall-har} and \ref{tb:overall-id}, respectively. For our method, the number of embedding $K$ decides the compression rate $\gamma$. The original communication burden is $3\times 114\times 500\times 4$ bytes per second (since one float number occupies 4 bytes), while our EfficientFi only transmits $K$ $\log_2 K$ numbers. For instance, with $K=256$, the quantized feature $E_d(x)$ consists of 256 embeddings, and our method transmits an ordered sequence of 256 integers varying in [1, 256]. The compression rate is calculated by $3\times 114\times 500\times 4\div 256\div 8=334$. In Table \ref{tb:overall-har} and \ref{tb:overall-id}, it is observed that our EfficientFi achieves satisfactory reconstruction error with a high compression rate for both human activity recognition and identification applications. The EfficientFi outperforms the compressive sensing methods (LASSO and BM3D-AMP), and the deep learning compression method (CSINet).

Since the comparative methods are only designed for compression, N/A indicates ``not applicable'' for the column of recognition accuracy. To look for a fair baseline, we use the same feature extractor $E$ and the classifier $G$ to form a normal classification network, namely ``EfficientFi-Encoder'' in Table \ref{tb:overall-har}. Compared to other two human activity recognition systems, our method outperforms E-eyes and CARM by 20\% around. Without quantization, the vanilla accuracy of our method is 98.3\%. After compression, it is observed that the accuracy drops as the compression rate increases, which is reasonable because the larger compression rate may hinder the feature discriminability. An interesting finding is that when the compression rate is low, the EfficientFi achieves even better performance than the vanilla EfficientFi-Encoder. Such phenomenon also happens for deep model compression using quantization \cite{polino2018model}. Overfitting can be prevented by simpler quantized feature representations that help our model attain better generalization capability. The unsupervised reconstruction also preserves more information for classification, achieving a marginal improvement. 

For person identification application, since the differences among CSI profiles of people gaits are quite small, the task is much more difficult. The WiWho and AutoID which leverage traditional machine learning and shapelet learning respectively for gait recognition only achieve 67.3\% and 77.6\%. Whereas, our encoder-only model can achieve 83.3\% accuracy. After compression, the EfficientFi can retain the accuracy of 83\% around. Better accuracy is achieved at a low compression rate of 66.8, which is similar to the better performance of activity recognition in Table \ref{tb:overall-id}. Then the recognition accuracy drops as the compression rate increases. We find that if we continue to increase the compression rate, i.e. reducing the number of embeddings to lower than 32, then the NMSE and accuracy will drop a lot. This demonstrates that $K=64$ should be the maximum compression rate of 1781.

In summary, the EfficientFi is able to achieve high fidelity with high compression rate while preserving high recognition accuracy for wireless human activity recognition and person identification.


\begin{table}[tp]
	\centering
	\caption{Performance Comparison on HAR}\label{tb:overall-har}
	\begin{tabular}{|l|c|c|c|}
		\toprule
		Method                   & $\gamma$ & NMSE (dB) & Accuracy (\%) \\ \midrule \midrule
		LASSO \cite{daubechies2004iterative}& 4        & -28.04     & N/A             \\ \midrule
		BM3D-AMP    \cite{mashhadi2020distributed} & 4        & -18.32     & N/A             \\ \midrule
		\multirow{4}{*}{CSINet \cite{wen2018deep}}  & 4        & -31.58    & N/A             \\
		& 16       & -28.63     & N/A             \\
		& 32       & -21.87     & N/A             \\
		& 64       & -18.24     & N/A             \\ \midrule
		E-eyes \cite{e-eyes}         & N/A        & N/A         & 73.3          \\
		CARM  \cite{carm}        & N/A        & N/A         & 79.5          \\
		EfficientFi-Encoder          & N/A        & N/A         & 98.3          \\ \midrule
		\multirow{5}{*}{EfficientFi} & 67      & -38.73    & 98.6          \\
		& 148     & -38.37    & 98.6          \\
		& 334     & -37.82    & 98.3          \\
		& 763     & -33.16    & 98.1          \\
		& 1781    & -28.75    & 98.1          \\ \bottomrule
	\end{tabular}
\end{table}

\begin{table}[tp]
	\centering
	\caption{Performance Comparison on Human-ID}\label{tb:overall-id}
	\begin{tabular}{|l|c|c|c|}
		\toprule
		Method                   & $\gamma$ & NMSE (dB) & Accuracy (\%) \\ \midrule \midrule
		LASSO \cite{daubechies2004iterative}& 4        & -27.37     & N/A             \\ \midrule
		BM3D-AMP    \cite{mashhadi2020distributed} & 4        & -17.87     & N/A             \\ \midrule
		\multirow{4}{*}{CSINet \cite{wen2018deep}}  & 4        & -29.18    & N/A             \\
		& 16       & -26.18     & N/A             \\
		& 32       & -20.40     & N/A             \\
		& 64       & -18.07     & N/A             \\ \midrule
		WiWho  \cite{wiwho}        & N/A        & N/A         & 67.3          \\
		AutoID \cite{zou2018identification}         & N/A        & N/A         & 77.6          \\
		EfficientFi-Encoder          & N/A        & N/A         & 83.3          \\ \midrule
		\multirow{5}{*}{EfficientFi} & 66.8      & -35.18    & 84.5          \\
		& 148.4     & -34.23    & 81.6          \\
		& 334.0     & -30.19    & 82.7          \\
		& 763.4     & -29.18    & 82.1          \\
		& 1781.3    & -27.70    & 83.3          \\ \midrule
		EfficientFi-Incremental          & N/A        & N/A         & 89.5          \\ 
		 \bottomrule
	\end{tabular}
\end{table}

\subsection{Incremental Learning via Reconstructed Data}
The reconstructed CSI can be further leveraged to fine-tune the classifier for better recognition performance. When the new CSI data is reconstructed in the server side, we can fine-tune the classifier using the high-confident samples. We use one half of the original testing data for fine-tuning and another half for evaluation. Setting the confidence threshold to 90\%, we obtain an improved accuracy of 89.5\% in Table III, denoted as EfficientFi-Incremental. This demonstrated that the reconstructed data can be beneficial to a boosted classifier for WiFi sensing.

\subsection{Hyperparameter Sensitivity}\label{sec:hyper}
\begin{figure}[t]
	\centering
	\includegraphics[width=1.0\linewidth]{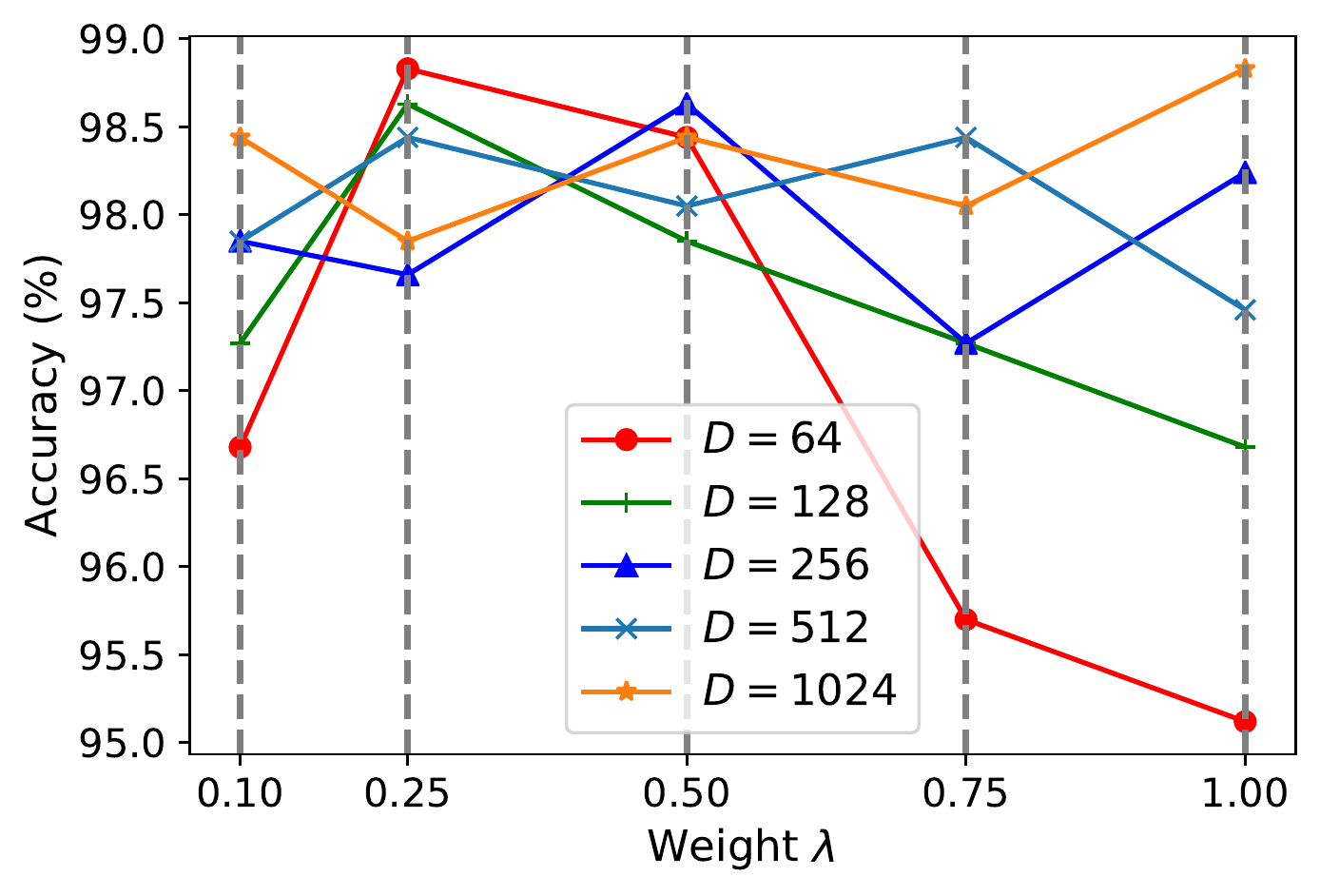}
	\caption{The hyper-parameter sensitivity in terms of the embedding dimension $D$ and the weight term $\lambda$ in Eq.(\ref{eq:main-eg}).}
	\label{fig:sensitivity}
\end{figure}

As our EfficientFi can always converge and achieve good performances with all selected code number $K$, here we investigate the embedding dimension $D$ and the weight $\lambda$ with the compression rate of $K=64$. When we choose the hyperparameters, we only focus on the recognition accuracy since their NMSE performances are acceptable for all settings. We summarize the results in Figure \ref{fig:sensitivity}. For different embedding dimension $D$, i.e. different line with five types of colors, it is seen that the variance of accuracies decreases as $D$ increases. With $D=64$, the accuracy can vary from 95\% to 98\%, while its range shrinks to 97.5\%-98.5\% with $D$ greater than 256. This is caused by the fact that larger embedding dimensions must contain more abundant information for representation learning, which could be more robust to other impact factors. Nevertheless, since a large $D$ contains more learnable parameters for EfficientFi, it can introduce more computation burdens and lead to larger training and testing cost. In our practical scenario, we attain a trade-off between such variance and computation complexity, and select $D=256$. Then we focus on the weight $\lambda$ that balances the CSI reconstruction and the classification task. Larger $\lambda$ can affect the performance of the classifier for the situations of $D\leqslant 128$, as shown at $\lambda=1.0$ when it is compared to those of $\lambda< 1.0$ in Figure \ref{fig:sensitivity}. It is noteworthy that $\lambda=0.5$ provides the most stable and satisfactory accuracy for our EfficientFi, and hence it is chosen to our default value of $\lambda$

\begin{figure*}[tbp]
	\centering
	\subfigure[Raw data]{\includegraphics[width=0.32\textwidth, angle=0]{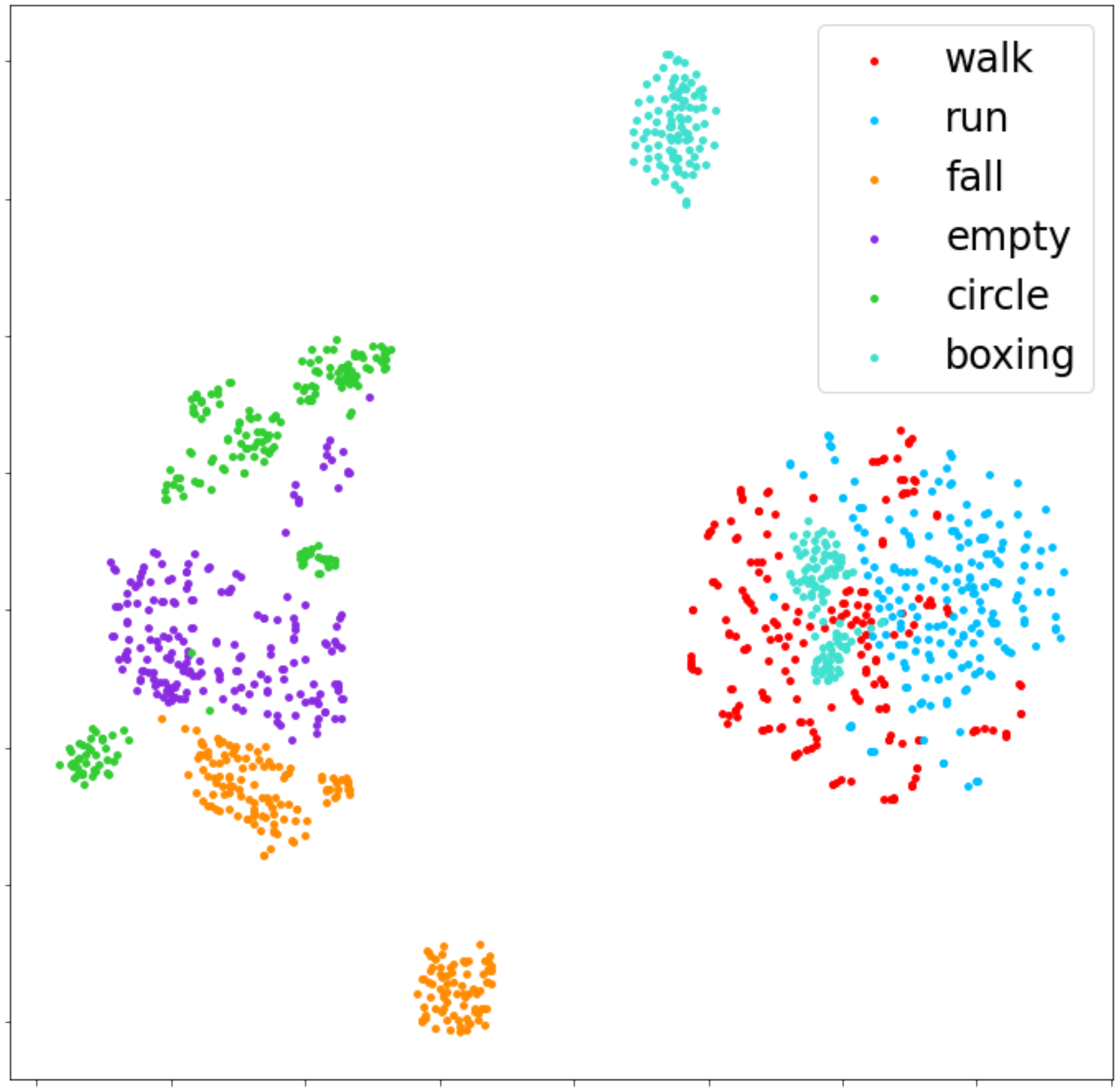}}
	\subfigure[Quantized Feature $E_d(x)$\label{fig:tsne-z}]{\includegraphics[width=0.32\textwidth, angle=0]{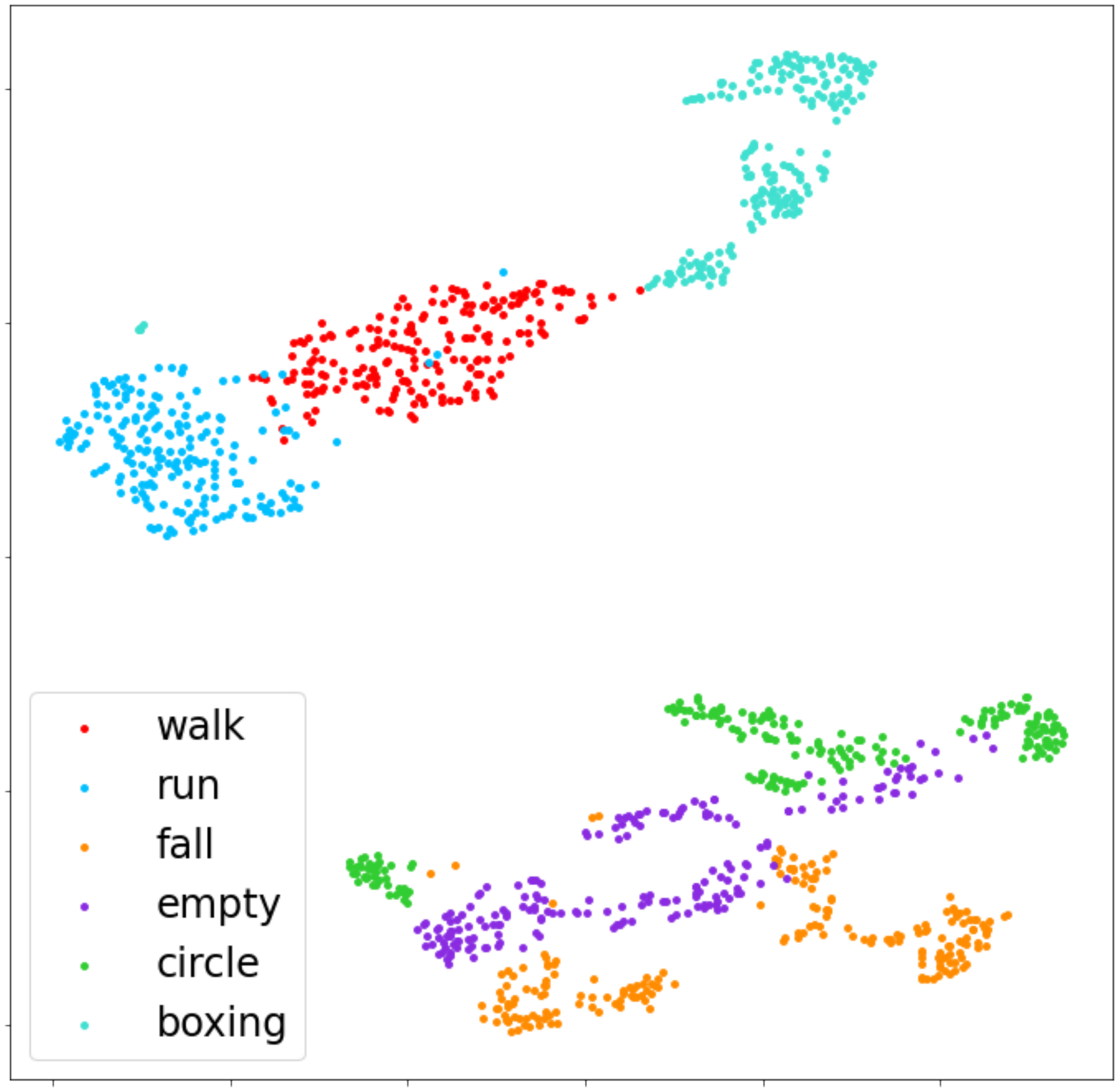}}
	\subfigure[Last Dense Layer\label{fig:tsne-c}]{\includegraphics[width=0.32\textwidth, angle=0]{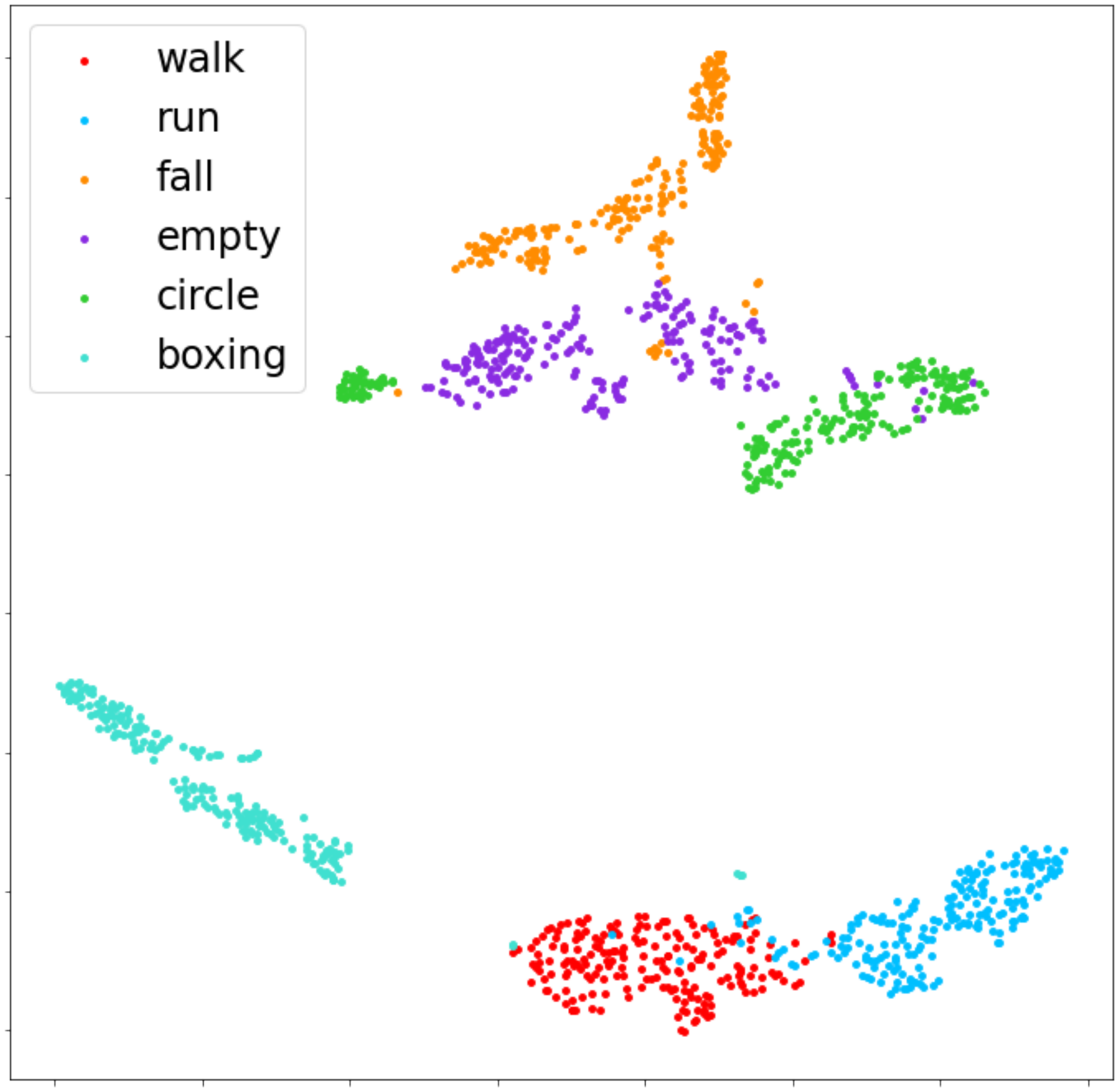}}
	\caption{\label{fig:tsne}The T-SNE embedding of the raw CSI data, the quantized feature $E_d(x)$, and the last dense layer of the classifier $G$.}
\end{figure*}

\subsection{Inference Time}\label{sec:rt}
We compare the inference time of the compression model, i.e. the encoder and quantization process, with existing methods. At the running setting on an NVIDIA RTX 2080Ti for deep models or Intel-I7 for LASSO, our method only spends 2.1ms to compress the CSI data of one second duration at 500Hz sampling rate. In comparison, the LASSO and BM3D-AMP take 251ms and 747ms, respectively. The CSINet shows better efficiency, and it takes 5.1ms for one timestamp of the CSI data. However, these methods only operate on a single CSI frame. The time cost will increase linearly as the sampling rate gets high. For example, a sampling rate of 500Hz can lead to $500\times 5.1$ms cost for the CSI-Net.


\subsection{T-SNE Visualization}\label{sec:tsne}
As shown in Figure \ref{fig:tsne}, the raw CSI shows that some activities are quite similar, such as \textit{walk} and \textit{run}. It is difficult to directly build a classifier for the raw data. For the proposed EfficientFi, the quantized features in the compressing space are shown in Figure \ref{fig:tsne-z} where all categories are separated to multiple discriminative regions (i.e. clusters). The feature space before the classifier in Figure \ref{fig:tsne-c} is quite similar to the compressed space. It is observed that the reconstructed feature still preserves the feature discriminability, and the classifier further aligns the latent space to form tighter clusters.


\section{Conclusion}\label{sec:conclusion}
In this paper, we propose a novel wireless sensing framework, EfficientFi, for large-scale WiFi-based sensing applications. The method firstly analyzes the limitations of existing approaches. To overcome these limitations, we design a quantized representation learning framework with a joint recognition learning scheme. The whole framework can be trained in an end-to-end and offline manner, and then it works online between the edge IoT sensing device and cloud server. Extensive experiments demonstrate that EfficientFi attains state-of-the-art performance compared to classic compressive sensing and deep compression methods. Moreover, the performance of classification task is well preserved after feature compression and transmission.

\bibliographystyle{IEEEtran}
\bibliography{egbib}

\end{document}